\documentclass[a4paper,11pt]{article}
\usepackage{aaskaiid,journal_names}
\usepackage{orcidlink}

\newcommand{\B}{$\vec{B}$}
\newcommand{\alfmm}{$\alpha_{\rm mm}$}

\newcommand{\asec}{$^{\prime\prime}$}

\newcommand{\Prot}{$\mathrm{P_{rot}}$}

\newcommand{\nup}{$\mathrm{\nu_p}$}
\newcommand{\nuB}{$\mathrm{\nu_B}$}

\newcommand{\teff}{T$_\mathrm{eff}$}
\newcommand{\Tbp}{T$_\mathrm{B}(\nu)$}

\title{Exploring Activity Across the Stellar Main Sequence with the Sun as a Benchmark}
\ShortTitle{Comparative studies of solar and stellar activity}

\author[1,2]{Atul Mohan\orcidlink{0000-0002-1571-7931}}
\ShortName{A. Mohan et al.} % shortened name list for header 
\author[3,4]{Stephen M. White\orcidlink{0000-0002-8574-8629}}
\author[5,6]{Sven Wedemeyer\orcidlink{0000-0002-5006-7540}}
\author[1,7]{Vladimir Airapetian\orcidlink{0000-0003-4452-0588}}
\affiliation[1]{Solar Physics Laboratory, NASA Goddard Space Flight Center, Greenbelt, MD, USA}
\emailAdd{atul.multiverse73@gmail.com}
\affiliation[2]{Department of Physics, The Catholic University of America, Washington, DC, USA}
\affiliation[3]{Space Vehicles Directorate, Air Force Research Laboratory, Kirtland AFB, NM, USA}
\affiliation[4]{Department of Physics and Astronomy, University of New Mexico, Albuquerque, NM 87106, USA}
\emailAdd{stephen.white.24@spaceforce.mil}
\affiliation[5]{Rosseland Centre for Solar Physics, University of Oslo, Oslo, N-0315, Norway}
\affiliation[6]{Institute of Theoretical Astrophysics, University of Oslo, Oslo, N-0315, Norway}
\affiliation[7]{American University, Washington, DC, USA}
\emailAdd{sven.wedemeyer@astro.uio.no}
\emailAdd{vladimir.airapetian-1@nasa.gov}

\abstract{
The active atmospheres of cool main-sequence stars (F-M type) often release a fraction of their stored magnetic energy, producing enhanced emissions (flares) across radio to X-ray wavelengths and associated space weather events like coronal mass ejections (CMEs) and energetic particle events (EPEs). Detailed imaging of active regions and CMEs, and in-situ EPE measurements are possible only in our Sun, making it a benchmark for stellar activity research. Multiwaveband solar imaging datasets let us define robust disk-integrated `Sun-as-a-star' diagnostics of active region and space weather, extendable to stellar datasets. Radio waveband provide diagnostics of particle acceleration, CMEs and EPEs, essential to model flare events and their space weather impacts. The Square Kilometre Array (SKA) telescopes will facilitate sub-second scale spectropolarimetric imaging of the solar corona across 0.05 - 15\,GHz, enabling detailed vertical tomographic studies of the active region across a range of coronal heights. Coupled with high energy instruments, the SKA telescopes will allow well-constrained modeling of large samples of diverse active phenomena and the defintion of robust Sun-as-a-star diagnostics of active region and space weather. Besides, the supreme sensitivity and angular resolution of the SKA telescopes will help detect quiescent and active emissions from several nearby stars. This chapter discusses the importance of comparative solar–stellar studies using Sun-as-a-star diagnostics in understanding activity and associated space weather conditions in stars across the cool main-sequence, and presents some research avenues that will benefit solar and stellar astrophysics.}

\begin{document}
\maketitle
\section{Introduction}\label{sec:intro}
Cool main-sequence stars (F–M type; cool stars) host most known habitable-zone exoplanets~\citep{Bashi20_occurrence_of_smallplanetsFGK}. They possess deep convective envelopes that power strong dynamos that generate strong and dynamic magnetic fields threading across the stellar atmospheric layers~\citep[see,][for a review]{Donati09_Rev_Bfield}.
%producing frequent flares, intense ultraviolet (UV) radiation, and hot coronae that emit soft X-rays~\citep[see,][for an overview]{2006astro.ph..9389G,Linsky16_Stellar_chromRev,benz17_flareRev}.
The active outer atmospheres in cool stars, including our Sun, are capable of producing major space weather events, including intense hour-long flares across the electromagnetic spectrum, especially in the ultraviolet to gamma ray bands, coronal mass ejections (CMEs), and { energetic particle events~\citep[EPEs;][]{kowalski24_Rev_flaremulti-band}}.
The properties of stellar convective envelopes and the emergent properties of magnetic field and activity depend on its rotation period (\Prot), effective photospheric temperature (\teff), and age~\citep[e.g.][]{Noyes84_RHK,Stepien94_Ro_activity_rot_relations,Donati09_Rev_Bfield,Vidotto14_B_Vs_age_n_rot}. 
Cool stars form high (`C') and low (`I') activity populations in the \Prot\,-\,age\,-\teff\ plane~\citep[][]{Barnes03_Rot_Vs_age_Vs_Activity,2014ApJ...789..101B}. The `C' branch has young ($<1$, Gyr), fast-rotating (\Prot$\lesssim$5\,days) stars and vise versa for the `I' branch. Stars migrate from `C' to `I' branch as they age, with lower \teff\ stars migrating earlier. 
Hence, it is essential to model the atmospheres and emergent activity of stars across \Prot\ - age - \teff\ plane.

{ The Sun–Earth system serves as a benchmark for stellar activity and space weather research, providing extensive multiwavelength imaging and in-situ energetic particle datasets from a wide range of space- and ground-based observatories.}
Data-driven modeling of solar active regions using these datasets can provide unprecedented insight into the drivers of major space weather events and associated active regions. 
Moreover, because the fundamental physical processes and the standard flare model of magnetic activity generally apply across cool stars~\citep{kowalski24_Rev_flaremulti-band}, the decades-long solar activity database provides a template library for developing robust solar disk-averaged (Sun-as-a-star) diagnostics of space weather events and active regions. These diagnostics can be extended to stellar studies that lack spatially resolved observations of active regions and direct measurements of space weather.

The energy-release events triggered by stellar activity are often observed as enhanced emissions or flares across the electromagnetic spectrum from long wavelength radio to $\gamma$-ray band.
%across the active stellar atmospheric layers, namely the chromopshere and the corona~\citep[e.g.][]{hirayam74_flarepromerupt_model_n_geom,shibata95_flarereconn_link,ash14_flareEnergetics_MagE,Ash15_flareEnergetics_therm_frmB}. 
%This heating and ionization of the plasma produce 
%Various ingredients of the standard flare model of magnetic activity, particularly the fundamental plasma physical framework, and the multiwaveband emission mechanisms generally apply to Sun-like stars~\citep[see,][for an overview]{kowalski24_Rev_flaremulti-band}. 
\begin{figure}[!htb]
 %\vspace{-0.3cm}
  \centering
  \includegraphics[width=0.94\textwidth,height=0.27\textheight]{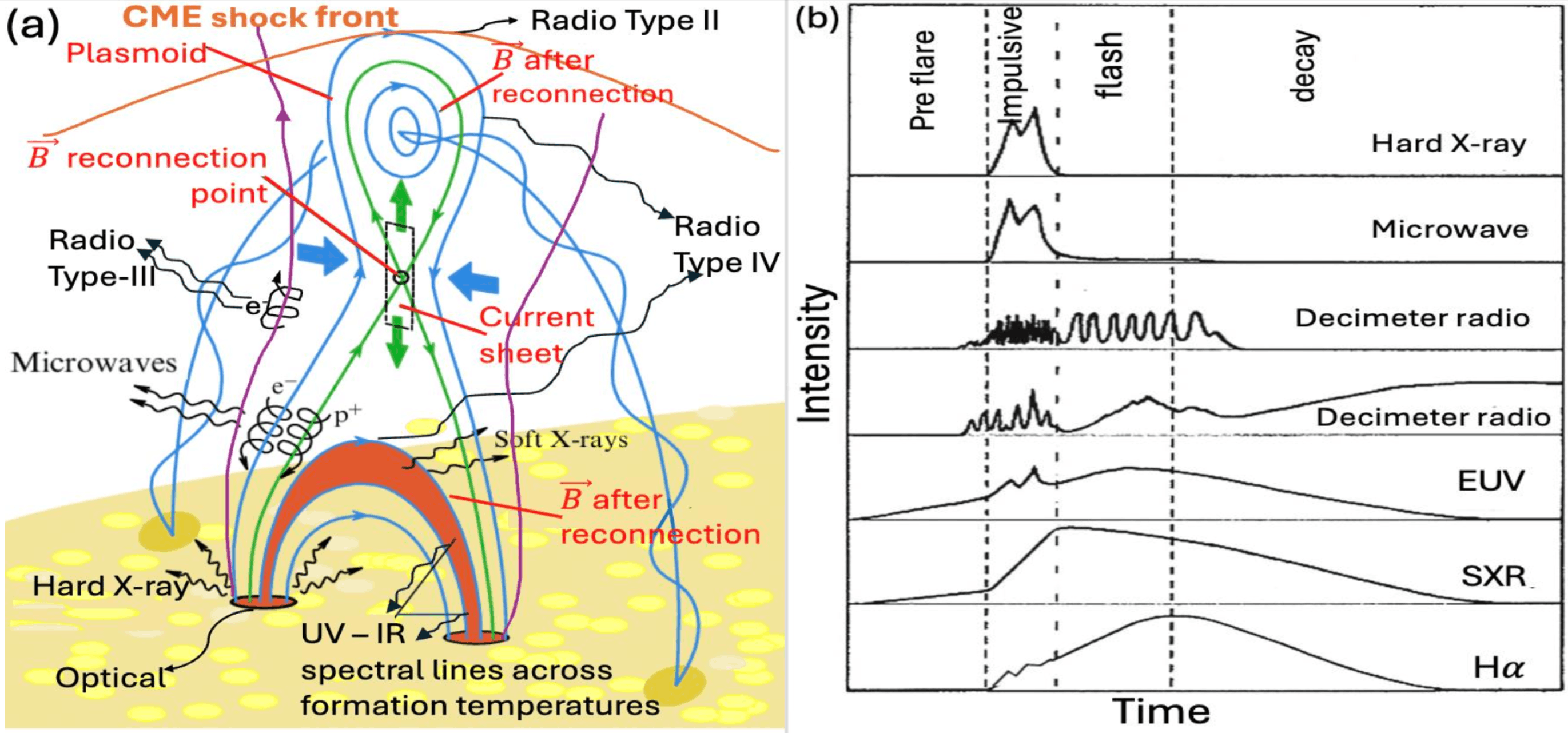}
\vspace{-0.3cm}
   \caption{{ Standard flare model. {\it (a):} Schematic of the standard model showing the various emission regions.} CME shock front (orange), open/closed field lines (violet/blue), and the reconnecting field line (green) are shown~\citep{lysenko19_xray-gamma_flares}. {\it (b): }Typical Sun-as-a-star multiwaveband light curves from the impulsive non-thermal and the gradual thermal phases, highlighting the Neupert effect (see the text for details).}
   \label{fig1}
   \vspace{-0.3cm}
\end{figure}
{ Figure \ref{fig1}a illustrates the standard flare model, in which magnetic reconnection releases a fraction of the stored free energy to accelerate the particles and subsequently heat the atmospheric layers, namely the chromopshere and the corona~\citep[e.g.][]{hirayam74_flarepromerupt_model_n_geom,shibata95_flarereconn_link,ash14_flareEnergetics_MagE,Ash15_flareEnergetics_therm_frmB}.} Strong flare events can also cause coronal mass ejections (CMEs) and produce major energetic particle events~\citep{gopal03_sun-geospace_conn,yashiro06_flare-CME_corr,gopal18_sgre_CME_link,kazachenko23_ErupConfflare_ML}.
{ The SKA telescopes operate in a frequency range that probes} both impulsive and sustained particle acceleration across the corona. They trace the propagation of energetic electrons and CMEs, thereby providing key space weather diagnostics. In particular, type III and type IV radio bursts are produced by accelerated electrons traveling along open field lines and contained in closed magnetic-field structures, respectively, when the electrons accelerated at CME-driven shocks generate type II bursts~\citep{McLean_solar_radiobook}.
Meanwhile, emissions at other wavelengths provide complementary information on flare evolution. Hard X-ray (HXR) probes the emission from the sudden impact of electron beams in the solar atmosphere. Soft X-rays (SXR) diagnose { subsequent} coronal heating, while the extreme ultraviolet (EUV) { continuum} and infrared-to-UV lines probe thermal and non-thermal plasma properties, and bulk flows across the chromosphere and the low corona.
Additionally, quasi-steady magnetic activity in active regions, even with weak or often undetectable heating signatures, can produce intense coherent type I radio bursts, offering diagnostics of persistent particle acceleration in the corona~\citep{iwai12_typeI_smallscaleEUVmagDyn_link,iwai14_typeI_finefeature_hists,Li17_typeI_smallEUVflare_link,Atul19_ARTB_microflare}. 
Additionally, solar observations close to 15\,GHz can trace the regions of the upper chromosphere associated with coronal holes that power the fast solar wind~\citep{gopal00_CHmagfield-17GHzTB_corr,akiyama13_17GHz-corhol}. 
%Several studies have also shown the importance 
%Millimeter (mm) emission primarily tracks chromospheric heating. 

Figure~\ref{fig1}b shows the typical solar-stellar flare light curves during various flare phases from the preflare to the decay phase. The impulsive phase at the start of the flare is a direct diagnostic of non-thermal particle acceleration { and the impact of energetic electron beams in the corona and chromosphere}, followed by the gradual heating phase, primarily seen in the high energy continuum emissions. The integral of impulsive phase emission often correlates well with the gradual phase thermal emission, as first noted by \cite{neupert68}, and henceforth is known as the Neupert effect. The Neupert effect highlights that the energy dissipation by the non-thermal particles produced during the impulsive phase gradually heats the corona.  

\subsection{Solar-stellar connection}
\begin{figure}[!htb]
 %\vspace{-0.3cm}
  \centering
  \includegraphics[width=0.94\textwidth,height=0.25\textheight]{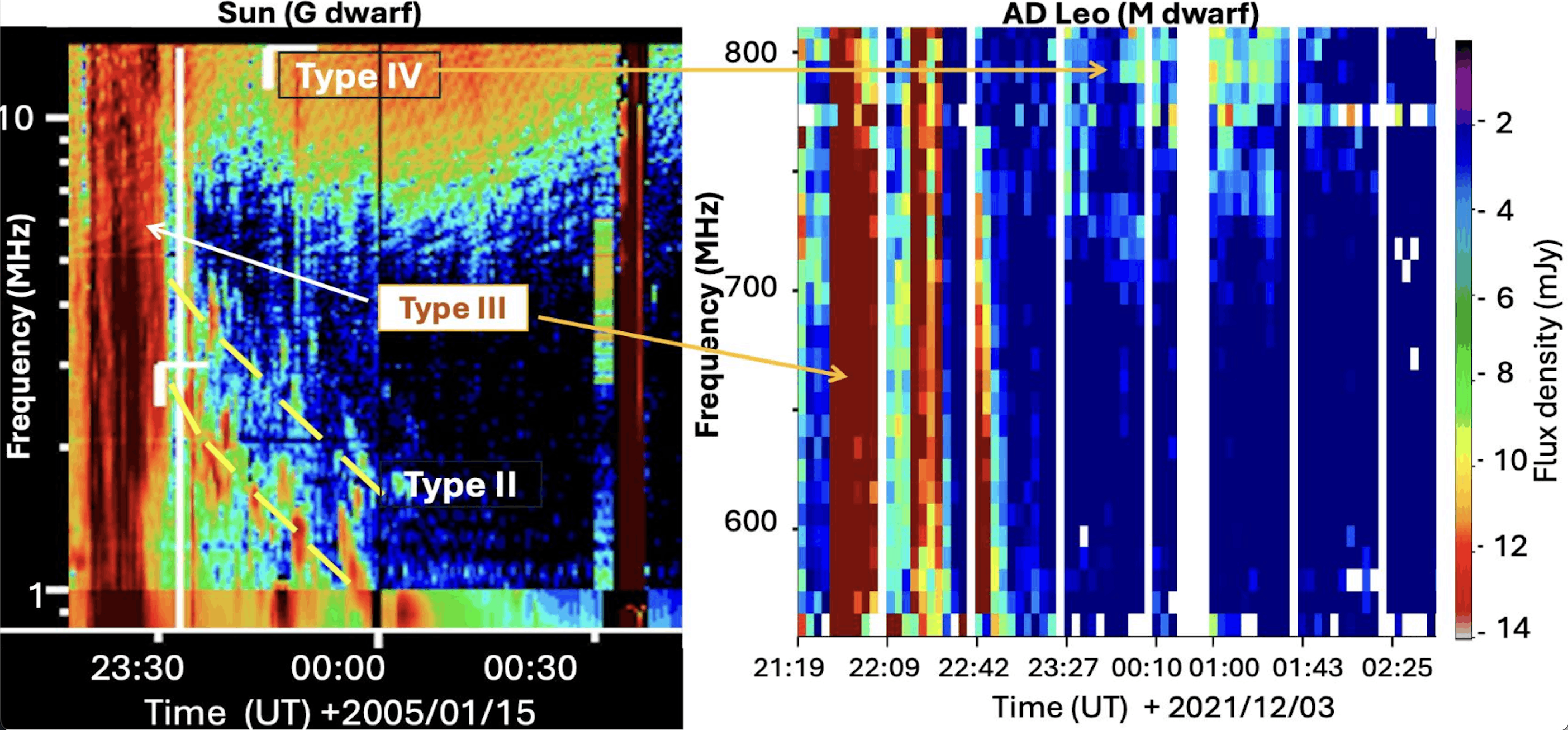}
\vspace{-0.3cm}
   \caption{{ Solar-stellar connection. Comparison of radio DS during a fast solar-CME (STOKES I) and during a burst reported in AD\,Leo (STOKES V), revealing the various radio burst types}~\cite{Atul24_ADLeotyIV,Atul24_typeIV-sun-ADLeo_comp}.
}
   \label{fig2}
   \vspace{-0.3cm}
\end{figure}

{ Several stellar flares demonstrate the Neupert effect~\citep[e.g.][]{Guedel96_NeupertEff_UVCet,Tristan23_Neupert-nonNeupert}. and show similarities in the timing of the onset and profiles of multiwavelength flares compared to the typical Sun-as-a-star template~\citep{kowalski24_Rev_flaremulti-band}. 
For instance, Fig.\ref{fig2} shows a typical solar radio burst dynamic spectrum (DS) during a fast solar CME with a DS from AD\,Leo during a major flare~\citep{Atul24_DHtypeIV,Atul24_ADLeotyIV}.}
%light curves in both solar and stellar events underscore their strong phenomenological correspondence~\citep{kowalski24_Rev_flaremulti-band}.
Beyond individual flare analogies, universal scaling relations such as the Güdel–Benz law~\citep[GBR;][]{guedel93_dM_GBZ}, linking radio and SXR luminosities across solar and stellar flares spanning several orders of magnitude~\citep{Benz94_GBZextend10ord_flares-sun-RSCVn,guedel95_GBR_FGgiants}, underscore the shared underlying physics of these phenomena. Additional studies have revealed consistent flux–flux correlations among spectral lines, continua, and surface magnetic flux for the Sun and solar-type stars~\citep[e.g.][]{pevtsov03_Xr-Rec_univcorr,toriumi22_flux-fluxcorr}. 
These results highlight the phenomenological correspondence between solar and stellar flares. 

{ Meanwhile, sensitive multiwaveband telescopes capable of observing stellar and solar flares with high sensitivity and spectro-temporal resolution are being commissioned. These sensitive facilities will enable detections of stellar flares over a large sample of stars with varying physical parameters, while also enabling detailed modeling of solar flares and derivation of robust space weather and active region diagnostics.
Extending solar-based diagnostics to a statistically significant sample of stellar events will help explore the long-term evolution of activity and associated space weather conditions around Sun-like stars, and contrast solar activity in the context of the broader main-sequence.}
%* Connection between solar-stellar flares and activity\\
%\hspace*{1cm}    >> Brief outline of standard flare model.\\
%\hspace*{1cm}    >> Radio imaging --> coronal tomography of non-thermal processes \\
%\hspace*{1cm}    >> Phenomenological correspondence~\citep{white04_solar-stellar_conn,Linsky16_Stellar_chromRev,kowalski24_Rev_flaremulti-band}\\

\subsubsection{Importance of Sun-as-a-star diagnostics in stellar space weather research}
{ Stellar astronomy is in a transformative phase with a new generation of sensitive multiwavelength observatories that are operational or are soon to be commissioned and can detect flares from large stellar samples. 
%Sensitive space and ground-based instruments are being planned in the radio to X-ray bands, including { the SKA telescopes}. 
%On the theoretical front, state-of-the-art MHD simulations can compute the impact of major space weather events on exoplanet atmospheres~\citep[see,][for a review]{Hazra25_rev}. 
However, there are no means yet or in the near future to directly detect and estimate the parameters of flare-associated space weather events or active regions.
%This critical knowledge gap hampers attempts to model not only the dynamics at active region scales but also the large-scale space weather and exoplanetary atmospheric impacts of major flares.
%On the contrary, the wealth of co-temporal multiwavelength imaging, magnetogram, and in-situ space weather measurement datasets facilitated by an array of space and ground-based instruments makes the Sun a benchmark for stellar activity and space weather studies. 
%Since fundamental physical processes and emission mechanisms during magnetic activity generally apply across cool stars, t
The vast solar database with detailed multiwaveband flare and associated in-situ EPE data will allow us explore robust Sun-as-a-star flare diagnostics that correlate well with the occurrence and properties of major space weather events and active regions.
A set of such robust Sun-as-a-star diagnostics in wavebands that are accessible to stellar research will facilitate the inference of space weather impacts of stellar flares, while also providing a parameter space for comparative solar-stellar activity research.}

Several studies have explored Sun-as-a-star diagnostics across multiple wavebands and their correlations with space weather and magnetic field properties. For instance, when it comes to CME diagnostics, flux dimming in EUV and SXR emissions provides a diagnostic of CME mass~\citep{hudson96_dimminginCME_Xflare,harrison03_cordimming_XUV,veronig21_Xraydim_sol-stel_data}, while spectral line Doppler shifts and line-width variations yield information on bulk plasma flows~\citep[e.g.][]{namekata24_EKdra_promerup}. 
In addition, statistical studies show that the stronger the SXR flare flux, the higher the chances of a CME~\citep{kazachenko23_ErupConfflare_ML}.
Nevertheless, the wealth of solar observations shows that many eruptive (CME-associated) flares lack one or more of these signatures, whereas some strong confined (CME-less) flares { exhibit some of }them~\citep{nitta17_CME-noXdim,veronig21_Xraydim_sol-stel_data}. 
In the case of SXR flares, about 10 - 20\% of solar flares belonging to the strongest energy class do not cause CMEs, while several flares belonging to an order of magnitude weaker class { are occasionally linked to CMEs}.
In this context, the relevance of radio diagnostics is notable, with $>$95\% of type II bursts and $>$81\% of type IV bursts associated with CMEs~\citep{Anshu21_CME_typeIV,anshu23_typeII_stats}. The association rates increase to nearly 100\% when the radio burst emission is observed to extend from decimetric to decameter-hectometric ranges~\citep{Gopal05_typeII-m-DH_SEPlink, Gopal11_PREconf,Miteva17_SEP-radburst_link,Atul24_DHtypeIV}.

\subsection{Radio diagnostics and their relevance}\label{sec:emiss_mech}
\begin{figure}[]
 %\vspace{-0.3cm}
  \centering
  \includegraphics[width=0.9\textwidth,height=0.34\textheight]{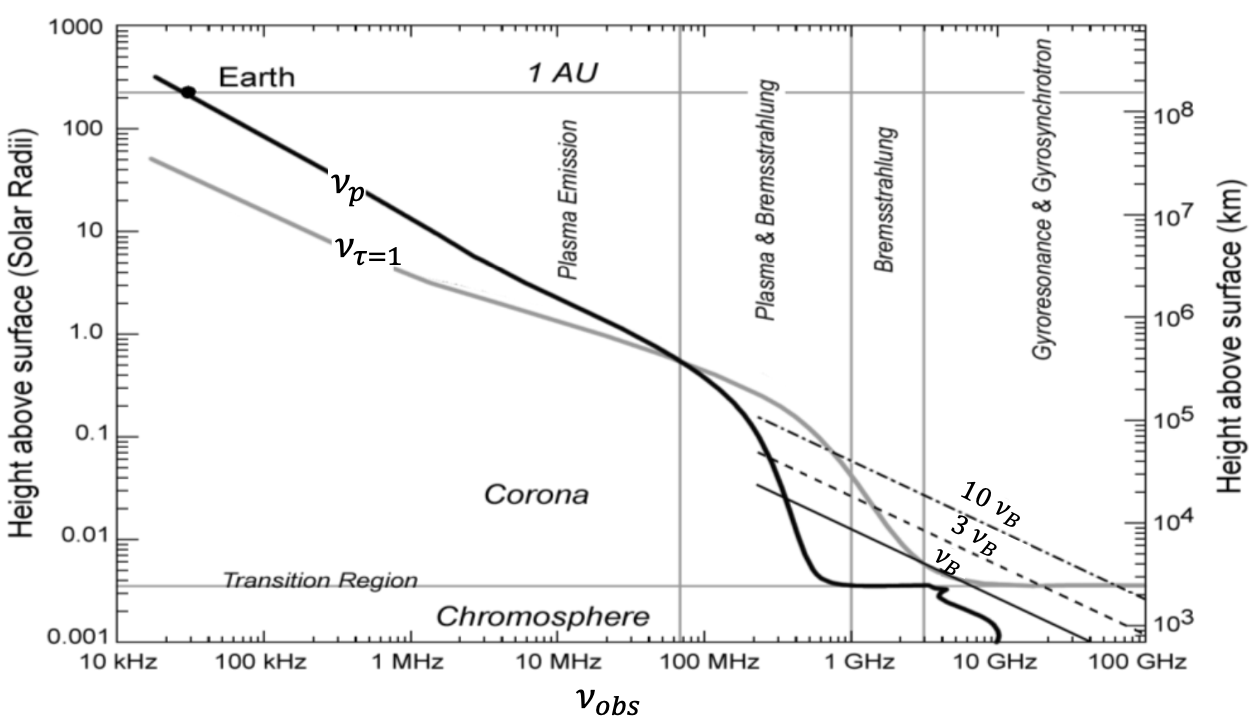}
\vspace{-0.3cm}
   \caption{{ Radio emission mechanisms. The profiles of \nup\ and \nuB\ are shown with the dominant emission mechanisms within various frequency ranges demarcated. The frequency at which optical depth becomes unity is shown as a function of height.} The density was computed based on VAL model B~\citep{VAL1981} and 5 times the \cite{1970AnTok..12...53S} model. Temperature profile was obtained from the VAL model to about 105 km, and was extended using a hydrostatic equilibrium model. Adapted from \cite{Gary04_plasmadiag}.}
   \label{fig_em}
   \vspace{-0.3cm}
\end{figure}

{ Solar-stellar radio} emission can be produced by five different mechanisms that are well understood and provide complementary diagnostics of physical conditions in the solar atmosphere:
\begin{itemize}
    \item { Bremsstrahlung}, also known as { free-free emission}, is due to the collisions of electrons with ions. At radio wavelengths, the opacity of this mechanism is dominated by thermal electrons and varies as $\mathrm{n^2 T^{-1.5} \nu^{-2}}$, where n, T, and $\nu$ denote the electron density, temperature, and emission frequency. The density dependence makes this a ubiquitous source of radio emission throughout the solar atmosphere. 
    %At high frequencies where it is optically thin, it produces a flat radio spectrum, which aids in its identification. At a given frequency, one sees down to the layer in the atmosphere at which that frequency is optically thick, providing temperature measurements thanks to the Rayleigh-Jeans limit.
    \item { Gyroresonance emission} results from the opacity of electrons orbiting magnetic field lines
at the gyrofrequency (\nuB\ = 2.8 $|$\B$_{Gauss}|$\,MHz). Field strengths of
order 100 - 1000\,G are common in solar active regions, and the corresponding values of \nuB\ lie in the microwave range. This mechanism provides high opacity purely due to the presence
of ambient thermal plasma and is therefore used to measure coronal magnetic fields~\citep{white97_gyroresonance_emiss_Rev}. This mechanism can produce high degrees of circular polarization depending
on whether the sense of polarization matches the sense of rotation of electrons about the
magnetic field.
    \item { Gyrosynchrotron emission} has the same basic physics as gyroresonance emission, but is produced by nonthermal high energy electrons accelerated in flares.
The radio spectrum can provide the physical characteristics of the radiating electrons. A typical flare has a spectral peak due to gyrosynchrotron at around 10 GHz, but in large solar
flares, gyrosynchrotron emission can be seen up to hundreds of GHz.
    \item { Plasma emission} is responsible for the well-known low-frequency solar radio bursts (Type I-V). It is a coherent mechanism where energetic electron beams efficiently excite electrostatic waves
at the Langmuir (or plasma) frequency, \nup\ = 9000 $\sqrt{{\rm n}}$\,Hz, and get converted to
propagating electromagnetic waves at the plasma frequency and its harmonic, 2\nup. %Electron beams have velocity distributions that are ideal for generating high levels of Langmuir waves,
Plasma emission is hence a powerful diagnostic of electron beams in
the solar atmosphere. However, absorption of plasma emission depends on a high power of electron density, and this limits its occurrence above several hundred MHz on the Sun.
    \item { Electron-cyclotron maser emission} (ECM) is another coherent mechanism that produces intense radio bursts in the frequency range around 1 GHz, immediately above
the frequency domain dominated by plasma emission. It is produced at $\nu$ = \nuB, and
is due to the structure in the 2-dimensional (perpendicular and parallel to the magnetic field) electron velocity
distribution that provides free energy for a relativistic resonance. 
%ECM powers strong planetary radio emission (e.g., auroral kilometric radiation from Earth,
Jovian decametric emission at Jupiter).
\end{itemize}
{ Figure~\ref{fig_em}, adapted from \cite{Gary04_plasmadiag}, summarizes the dominant emission mechanisms with various spectral bands and the variation of \nup\ and \nuB\ with height based on standard 1D atmospheric models. The plasma emission mechanism powers active solar/stellar emissions in { the operational band (0.05–0.35 \,GHz) of the SKA-Low telescope}. 
The supra-thermal electron beams accelerated at reconnection sites and CME-driven shocks produce various types of radio bursts with characteristic dynamic spectral variability (type I - IV; see Sect.~\ref{sec:intro}), and flux densities exceeding the quiescent emission by several orders of magnitude. 
These burst types form excellent diagnostics of active regions, CMEs, and solar energetic particle (SEP) events~\citep{Gopal05_typeII-m-DH_SEPlink,Miteva17_SEP-radburst_link}. Almost all type II bursts and more than 85\% of type IV bursts are CME-associated~\citep{gopal00_typeIIRev,Gopal11_PREconf, anshu23_typeII_stats,Hillaris16_typeIV,Atul24_DHtypeIV}.}
About 75\% of type IV bursts accompany fast, wide CMEs, which are highly geo-effective, often producing strong SEPs~\citep{Miteva17_SEP-radburst_link,Anshu21_C24typIV_CME_corr,Atul24_DHtypeIV}. \citet{atul24_modGBR} further derived relations connecting active-region and CME parameters to Sun-as-a-star type II burst observables and SXR flux. Type III bursts trace electrons propagating along open magnetic field lines, providing direct diagnostics of impulsive SEPs~\citep{2002JGRA..107.1315C,2015ApJ...809..105W,Miteva17_SEP-radburst_link}.
%All these bursts are primarily powered by coherent emission processes with orders of magnitude higher brightness temperatures than the quasi-steady solar disk emission, making disk-integrated flux densities almost entirely event-driven~\citep{ginzburg1958,Tsytovich69,melrose1972,Gary04_plasmadiag,white04_solar-stellar_conn}. 

Meanwhile, at higher frequencies covered by the { SKA-Mid telescope }(0.35–15\,GHz), both coherent (ECM) and incoherent mechanisms (gyrosynchrotron, gyroresonance) power the emission from flare-accelerated electrons~\citep[see][]{gary2001,guedel02_Rev_stellarRadioEmiss}. The emission mechanisms can be inferred from the dynamic spectral evolution and polarization. Coherent bursts, forming near harmonics of the local \nup\ or \nuB\ enable direct inference of local plasma parameters. Modeling incoherent emission, on the other hand, constrains the electron energy distribution and magnetic field strength in active regions, particularly when combined with spectro polarimetric diagnostics~\citep[e.g.][]{Dulk85_FlareradioEmissMech,1996ApJ...464..965G,2000ApJ...534L.203W,Lee03_multiband_mcrwavestudy,Gary18,alissandrakis21_MicrwaveRev,2023ApJ...948..121T,2025ApJ...991..186K}.

Several solar studies have also reported quasi-periodic pulsations (QPPs) and fine structures in 0.05 - 15\,GHz dynamic spectra, linking them to local plasma or MHD wave processes, density perturbations, { and particle acceleration scales} in active regions~\citep[see,][for a review]{Asch_QPPtheoryRev1987,Arzner1999,2009SSRv..149..119N,2023Univ....9..442A}. 
{ The SKA precursors and pathfinders are already making major progress in this field, which will also benefit the exploration of robust Sun-as-a-star metrics and insights extendable to stellar flare studies~\citep[e.g.][]{kontar2017,sharykin2018_LOFAR_dnn_withtypIIIb,Atul19_typeIIIQPP_turb,atul21_structQPPs,mondal21_longdurARQPPs,Atul21_dNN_vsht}. 
%We refer the readers to the relevant chapters in this book for a detailed overview of these topics. 
%Modeling the physics of these fine spectro-temporal variability during active solar emissions and deriving robust Sun-as-a-star physical diagnostics of the active region will be useful for stellar flare studies.
Besides their physical relevance, radio bursts frequently outshine the integrated quiescent solar or stellar disk emission by several orders of magnitude, making them excellent Sun-as-a-star activity diagnostics detectable from sources at far greater distances than quiescent stellar emission.} %Collectively, these aspects underscore the unique Sun-as-a-star diagnostic power of wideband radio observations for probing major space weather events and the properties of associated active regions.

\subsubsection{Tomographic exploration with wideband radio imaging spectroscopy}
\begin{figure}[!htb]
 %\vspace{-0.3cm}
  \centering
  \includegraphics[width=0.6\textwidth,height=0.22\textheight]{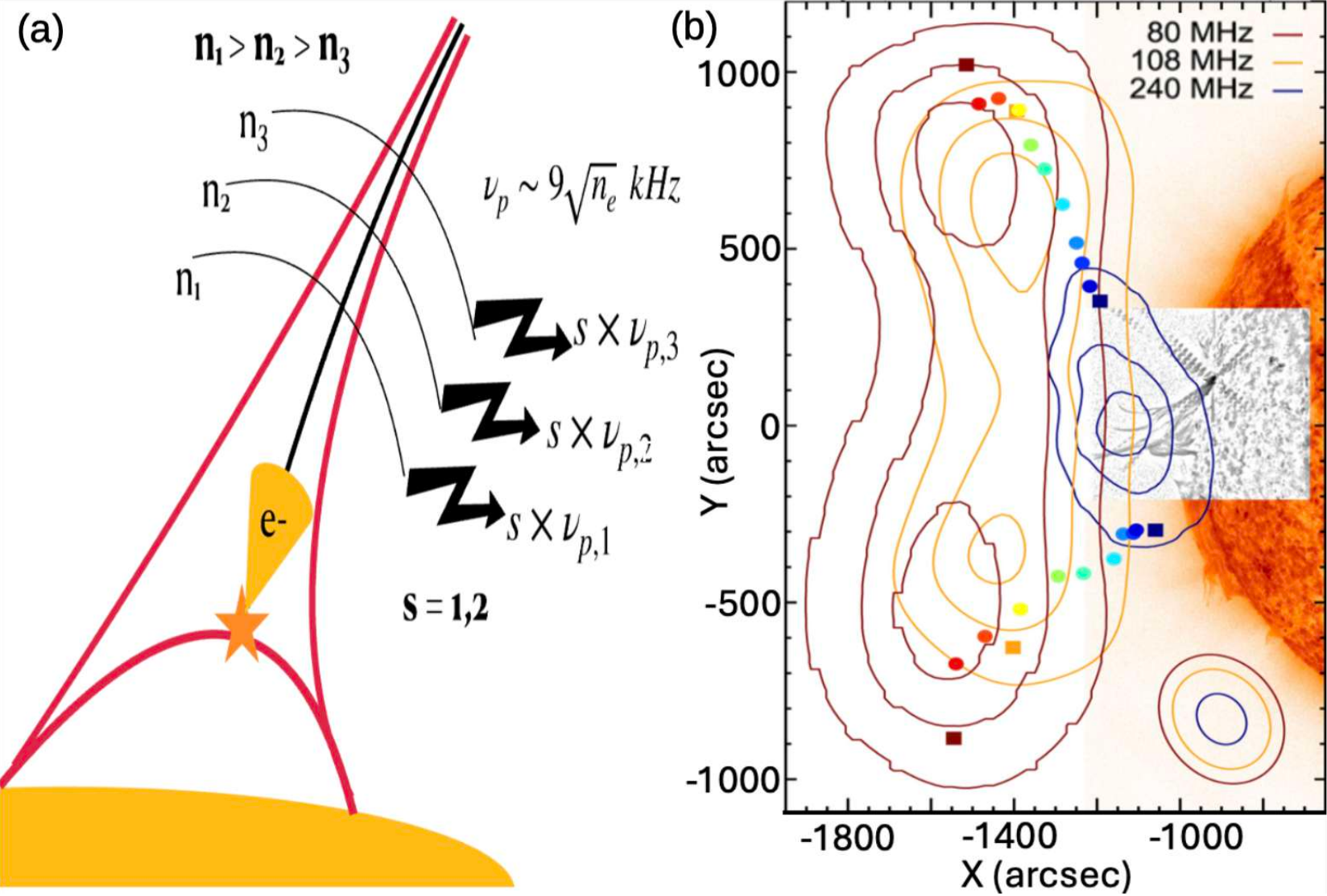}
\vspace{-0.3cm}
   \caption{{ Tomographic exploration using spectroscopic radio imaging.}
   %{\it (a): }Radial \nup(r) based on \cite{newkirk61_Nmodel} coronal electron density (n) model and \nuB(r) assuming a mean magnetic field model for active regions by \cite{Asch99_activereg_loopAnalyticalmodel} with varying footpoint magnetic field strengths (B$_{\rm foot}$). 
   {\it (a): }{ Illustration of coherent radio bursts triggered at varying density ((n$_{\rm i}$; i$\in [1,3]$)) regions, located at different heights along the tragectory of an accelerated electron beam. The burst emission occurs at the harmonics (s) of the local \nup.} {\it (b):} { Tomographic mapping of a solar active region using spectroscopic images within 80 - 240\,MHz range from the MWA. For each observing frequency, contours at 20, 50, and 80 \% of the peak intensity are shown. Colors from red to blue indicate increasing frequency, and marker separation denotes the source size at each frequency.
   The image contours at various frequencies collectively trace the vertical magnetic structure transporting the flare-accelerated electrons.}~\citep{patrick17_MWAQSL}.}
   \label{fig2}
   \vspace{-0.3cm}
\end{figure}
The different types of stellar active phenomena energise plasma and accelerate particles across coronal heights, often evolving into major space weather events. 
{ Hence, to study the evolution and space weather impacts of an active event, we require a robust tomographic exploration strategy that probes the event-associated dynamics across coronal heights.
A strategy to define a tomographic tool is to identify a set of spectral bands sensitive to specific thermal and non-thermal processes within narrow height ranges across the corona to interplanetary space.}

%{ Consider the SXR continuum and spectral lines formed in the corona. The SXR emission is optically thin, providing thermal diagnostics averaged over a very wide height range along the line-of-sight. }
{ Spectral line inversions using simultaneous multi-line data are often used to infer the heights and dynamics of plasma and energetic particles across the chromosphere to the corona. However, various non-local thermodynamic equilibrium (non-LTE) processes and propagation effects introduce degeneracy in the multi-line inversion procedure, making it difficult to accurately infer emission heights~\citep[e.g.][]{2017A&A...605A..53M,2017ApJ...836...35J,2017PhRvL.118o5101K,2022ApJ...926..223Z}.
Besides, the} time resolution obtained in multi-line spectral inversion studies is of the order of $\sim$ 10\,s to min even for stellar flares, making it difficult to trace seconds-scale non-equilibrium dynamics, known to exist from high cadence radio and X-ray observations~\citep[e.g.][]{Dennis85_HXR_highcadence_bursts,Osten2008,Endo10_HXRFlareCatalog,Hilaire13_typIIIstats}.
{ Meanwhile, the optically thin coronal plasma continuum in the SXR band provides thermal diagnostics averaged over a very wide height range along the line-of-sight.}

%In contrast,
{ %The characteristic local \nup\ and \nuB\ in the coronae of cool stars fall in the radio band explored by the SKA. Figure.~\ref{fig2}a shows radial \nup\ and \nuB\ profiles for the Sun based on typical 1D models.
%wideband sub-second radio imaging and non-imaging studies with the state-of-the-art and planned facilities like the SKA, enable robust tomographic exploration of highly variable non-thermal activity in both the sun and stars. 
%This is because of the emission mechanism in the radio band and the high emission brightness that outshines the entire solar/stellar disk even during weak particle acceleration processes~\citep{McLean_solar_radiobook, guedel02_Rev_stellarRadioEmiss,white2007_radiobursts}.
In contrast, coherent radio bursts associated with active events form at the harmonics of the local \nup\ or \nuB\, both of which vary with height in the corona (see Fig.~\ref{fig_em}). This property makes spectroscopic radio imaging across wide bandwidths at sub-second timescales an excellent tool to perform  tomographic studies of the onset and propagation of energetic particleas and their progenetors (e.g., CME) across corona and into the interplanetary space causing space weather impacts.
Figure.~\ref{fig2}a illustrates this idea using a cartoon of an accelerated electron beam propagating along an open magnetic field structure generating coherent bursts at varying frequencies as a function of height.
%Hence, spectroscopic radio imaging across wide operational bandwidths at sub-second cadence, facilitated by modern interferometers, is a powerful tool to track the dynamics of energetic particles and their progenetors (e.g., CME-shock) across coronal heights.
Figure.~\ref{fig2}b shows snapshot spectroscopic images from MWA within 80 -240\,MHz. The image contours clearly map the vertical structure of the active region magnetic field.
%Besides, the high emission brightness of radio bursts that outshine the intergrated flux density of the quiescent solar/stellar disk by several orders of magnitude even during weak activity, make them excellent activivity diagnostics~\citep{McLean_solar_radiobook, guedel02_Rev_stellarRadioEmiss,white2007_radiobursts}. This is particularly beneficial in stellar activity studies that lack spatially-resolved observations. 
Hence, modeling the dynamic spectral variability of the stellar coherent bursts by extending insights and diagnostics from solar studies can help constrain the evolution of energetic particles, instabilities, and their progenetors during active phenomena along the line-of-sight axis, providing valuable insights into the associated space weather impacts.}
\subsection{Radio instrumentation requirements for comparative solar-stellar studies}\label{sec:req_sol-stel}
%The solar-stellar studies primarily { address the questions of how the solar atmospheric dynamics and activity compare with other cool stars}, how a typical Sun-like star evolves over its lifetime, and whether the Sun is just another typical G-dwarf despite being the only known biosphere host. 
{ The interdisciplinary field of comparative solar-stellar research relies on a Sun-as-a-star analysis framework, which enables robust quantitative comparisons between solar and stellar phenomena.}
Developing a robust Sun-as-a-star framework requires characterizing the solar emission variability across dimensions of { flux density}, frequency, time, and polarization in the image plane to identify the sources, {understand the drivers of magnetic activity}, and define meaningful multi-dimensional diagnostics sensitive to major space weather events and active region properties despite spatial averaging of the data.

The dynamic spectral studies of the Sun and magnetically active stars have reported signatures of polarized emission variability at sub-MHz and sub-second scales.
{ Hence, the identification and tracking of flaring regions require wideband spectro-polarimetric imaging at sub-MHz and sub-second scales with sub-arcmin angular resolution, necessary to resolve active regions.
Additionally, there can be multiple co-temporal active events on the solar disk capable of producing radio bursts that are brighter than the event of interest by several orders of magnitude.
So, robust tracking of the evolution of a single event in the image plane requires high fidelity and high dynamic range imaging capability, which directly translates to the need for dense instantaneous {\it uv}-plane coverage enabled by a compact core large-N array configuration.}

{ On the stellar side, we require the capability to perform polarimetric imaging of stellar fields with high spectro-temporal resolution comparable to solar observations. To study activity across the cool main sequence, wideband, long-term monitoring datasets are required for stars spanning a range of \Prot, age, and \teff\ values. For meaningful comparisons with the Sun, target stars should be either isolated or members of non-interacting binary systems with a typical separation of $\gtrsim$20 AU like the $\alpha$\,Cen AB. Telescopes with high sensitivity and angular resolution of the order of a few arcseconds to sub-arcseconds are therefore essential for generating large samples of such stars by scanning sources within several tens to a few hundred parsecs.}
%high sensitivity enables detections at larger distances, while high angular resolution allows non-interacting binaries to be spatially resolved. Typical wide-binary stars are separated by $\gtrsim$20 AU, corresponding to angular scales of a few arcseconds to sub-arcseconds at distances of a few tens of parsecs.}

To quantitatively assess the sensitivity requirements, let us consider the Sun and AD\,Leo, { an active star located at 5\,pc}.
The typical solar quasi-steady flux can vary during its activity cycle within 50 - 100\,SFU~\citep{white04_solar-stellar_conn,2011SoPh..273..309S,Shimojo17_1950-2015_GHz-TSIVariab} { in the 0.05 - 0.35 GHz range (SKA-Low's operational band), while in 0.35 - 15.4\,GHz range (SKA-Mid's operational band), the flux can be about a few to 60\,SFU}~\citep{McLean_solar_radiobook,Div17_SolSpec}.
If Sun were located at 1 pc, its disk‐integrated quiescent flux would be $\sim$0.5 -30\,$\mu$Jy in the { SKA-Low's} and $\sim$ 25 - 50$\mu$Jy in the { SKA-Mid's operational bands}. 
Additionally, solar radio bursts can be 1 - 2  or 3 - 6 orders of magnitude brighter than the quiescent emission~\citep[see,][]{Hilaire13_typIIIstats,Nindos00_SXR_gyroresonance_Sunspots, Reid2014}.
{ Likewise, if AD Leo were located at 1\,pc, it would have a quasi-steady flux of $\sim$50-100\,mJy} in the 0.5 - 1\,GHz range~\citep{villadsen19_Cohbursts_but_notypeII,Atul24_ADLeotyIV}. 
%Such intense quiescent emission is seen amongst flare stars, hinting at a quasi-steady particle acceleration at active regions.
{ During strong flares, AD\,Leo can be 10 – 1000 times brighter than the quiescent level}~\citep{White89_VLA1GHz_starcatalog,Osten2006, villadsen19_Cohbursts_but_notypeII}.
%In comparative solar-stellar studies, it is important to focus on isolated or non-interacting binary stars. Non-interacting binaries with a typical separation $\gtrsim$20\,AU, corresponding to the separation scale in the $\alpha$Cen\,AB system. It is important that the angular resolution and sensitivity of the instrument can enable the detection and resolution of wide cool star binaries out to large distances to have a sufficiently large sample of non-interacting cool dwarfs.

\section{Advancements with the SKA telescopes}
The { SKA telescopes} will 
%probe the 0.05 - 15\,GHz spectral range that traces non-thermal particle acceleration processes (see, Fig.~\ref{fig_em} \& \ref{fig2}) across a wide range of coronal heights, providing diagnostics of plasma and magnetic field variability, and associated CMEs and EPEs. The arrays will 
{ facilitate sensitive sub-second sub-MHz snapshot spectro-polarimetric solar and stellar imaging observations in the 0.05 - 15\,GHz range.}
The large-N architecture of the { SKA telescopes} will enable high-fidelity imaging required for solar activity studies. The { telescopes} will offer significantly higher sensitivity across their operational bands than current state-of-the-art facilities, owing to their higher collecting area (A$_e$) and lower system temperature (T$_{sys}$) values. { Besides, owing to their long baselines, the angular resolution of the SKA telescopes will be an order of magnitude finer than the leading telescopes in the 0.05-15\,GHz range.}
%\paragraph{Comparison:}
%Sun at 1 pc : 4uJy (2MK); 4mJy (2$\times$10$^9$K) [200 MHz]\\
%ADLeo at 5pc $\sim$2mJy in quasi-steady state. So 1pc: $\sim$50mJy [quiescent] [$\sim$0.5 - 1 GHz]\\
%Flares $\sim$ 10  - 1000x.

\subsection{SKA-Mid in AA4 and AA* configurations}
The SKA-Mid { telescope}, planned to operate { over 0.35 - 15.4\,GHz, will enable high dynamic range imaging at about 0.14\,s and 13\,kHz resolution, with an angular resolution of about 0.23$^{\prime\prime}$ $\times$ 0.2$^{\prime\prime}$ in} the AA4 configuration at 1.4\,GHz. 
Meanwhile, the resolution for the AA* configuration is expected to be about 0.78\asec $\times$ 0.29\asec.
This is a significant improvement compared to the few arcsec resolution offered by the solar imaging mode of the Jansky Very Large Array (JVLA)\footnote{\href{https://science.nrao.edu/facilities/vla/docs/manuals/obsguide/modes/solar}{https://science.nrao.edu/facilities/vla/docs/manuals/obsguide/modes/solar}}and the 0.25$^{\prime\prime}$ resolution offered by the most extended JVLA configuration for general astronomical observations.
The sensitivity (measured as A$_e$/T$_{sys}$) of the AA* (AA4) array at 1.4\,GHz is expected to be about 6(8) times better than { MeerKAT~\citep{Jonas16_MeerKAT}, which is an SKA precursor.}
The AA*(AA4) array { would help achieve} a root mean square (RMS) noise level of 15.8(10.2)\,$\mu$Jy/beam with 1\,h and 0.81\,GHz averaging in Band 2, centered at 1.4\,GHz. 
This sensitivity will let us detect quiescent emission from { stars like AD\,Leo out to 25 (30)\,pc using the AA*(AA4) array} configuration, with 5$\sigma$ detection significance.
{ On the contrary, significant detections of radio bursts from stars at similar distances will be possible with temporal averaging of 0.1–1 s and spectral averaging of 0.1–1 MHz.} The strongest class of coherent bursts could be detected with similar spectro-temporal averaging by AA* and AA4, from stars at distances of $\sim$0.1\,kpc.
{ At 0.1\,kpc, 20\,AU corresponds to 0.2\asec, which will be barely resolved by the AA4 configuration, while AA* will resolve all binaries with a separation $>$80\,AU and located at distances up to 0.1\,kpc.}

\subsection{SKA-Low in AA4 and AA* configurations}
The sensitivity of the { SKA-Low telescope} (operational band: 0.05 - 0.35\,GHz) at 70 (200) MHz will be $\sim$120 (400) m$^2$/K in AA* and 200 (600)  m$^2$/K in AA4 configurations.
At 70 (200)\,MHz, the AA* array will be $\sim$ 15 (5) and AA4 about 28 (7.5) times better in sensitivity than the SKA pathfinder, LOw Frequency ARray~\citep[LOFAR;][]{vanHaarlem13_LOFAR}.
%Compared to the state-of-the-art instrument in the 0.2-0.5\,MHz range, the upgraded Giant Meterwave Radio Telescope~\citep [uGMRT][]{}, 
{ In the 0.2-0.5\,MHz range, the sensitivity of the SKA-Low telescope will be higher than the state-of-the-art instrument, the upgraded Giant Meterwave Radio Telescope~\citep [uGMRT][]{gupta17_uGMRT}, by a few factors.}
The SKA-Low telescope will achieve an RMS noise level of 81 (132) $\mu$Jy with 1\,h integration at 200\,MHz in AA4 (AA*) configuration, with an angular resolution of $\sim$3.89\asec.

{ The high sensitivity and large fields of view of the SKA telescopes} directly translate to orders of magnitude higher survey speeds in the radio band, compared to the current state-of-the art facilities. 
%The combination of higher sensitivity and angular resolution of the SKA { telescopes} will significantly increase the number of stellar flares observed. 
{ The faster survey speeds coupled with high sensitivity will help generate large samples of stellar flares with significantly less telescope time.} Additionally, the imaging high dynamic range will advance solar activity and Sun-as-a-star research.
All sensitivity estimates { presented here} are based on the { Square Kilometre Array Observatory's} sensitivity calculator\footnote{\url{https://sensitivity-calculator.skao.int/}} and the data released by the observatory online\footnote{\url{https://www.skao.int/en/science-users/118/ska-telescope-specifications}}.

\section{Research avenues}
A combined solar-stellar flare and space weather analysis framework is essential to infer not only the space weather relevance of flares across { the} stellar main-sequence but also to compare solar atmospheric structure and activity against other main-sequence stars across the \Prot-\teff-age.
Comparative solar stellar studies focusing on Sun-like G dwarfs of varying ages can also provide a picture of solar activity evolution across ages and test the 
%These studies will also aim to test the extent of 
applicability of { the standard flare model and emission mechanisms across a wide range of flares in stars with different atmospheric plasma and magnetic field parameters.} 
%Certain observations in active `C' branch stars, including young solar analogs, have demonstrated signatures of departures from the expected Fig.~\ref{fig_em}b template.
The following subsections will discuss some of the solar-stellar research avenues that will benefit from the { SKA telescopes}.

\subsection{Sun-as-a-star classification scheme for solar-stellar activity}\label{pspreds}
As discussed in Sect.~\ref{sec:emiss_mech}, there are multiple emission processes in the broader radio wavelength range, and the dominance of the emission process depends closely on the local { plasma and magnetic field parameters and their variability} scales~\citep{gary2001,Zhang24_ECMconditions}.
{ Hence, depending on atmospheric physical parameters, different emission mechanisms may dominate within a given observing frequency range for different stars}. 
There have been reports of mixed emission mechanisms and deviations from the general picture presented in Fig.~\ref{fig_em} { even for solar radio bursts}~\citep[e.g.][]{morosan16_ECM_conditionsCorona, Carley19_lossconeQPP, Morosan19_typeIV_emissmech}.
\begin{figure}[!htb]
 \vspace{-0.2cm}
  \centering
  \includegraphics[width=0.94\textwidth,height=0.56\textheight]{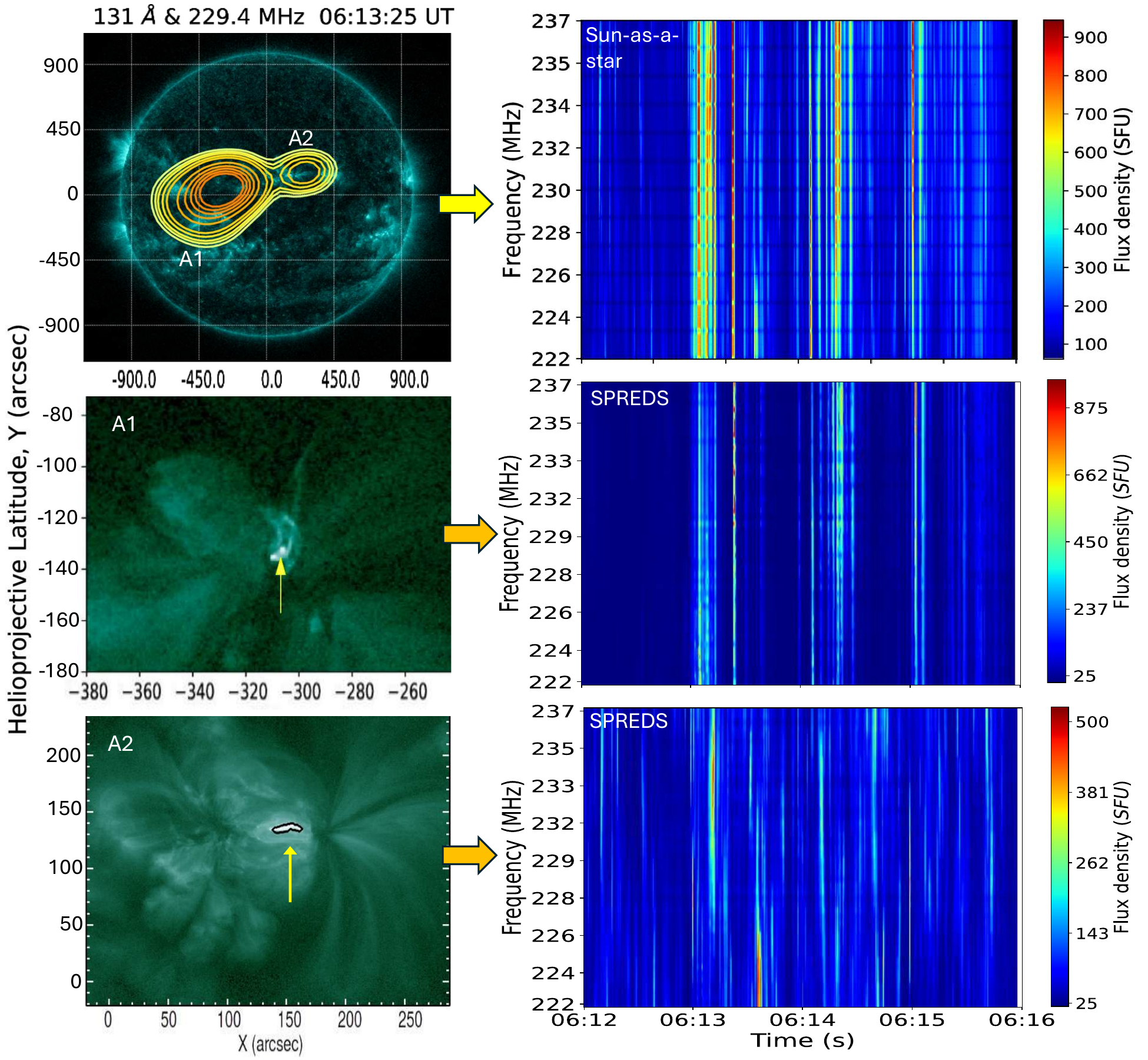}
\vspace{-0.3cm}
   \caption{{\it Top:} A sample radio contour map of the Sun from Nov 3, 2014, overlaid on co-temporal EUV image at 94\AA\ from SDO/AIA instrument. Radio contours are marked at 0.5, 0.7, 0.9, 1.3, 1.8, 3.6, 5.4, 7.2, 9, and 11 $\times 10^8$ K to highlight the co-temporal radio burst sources, A1 and A2. A Sun-as-a-star/total flux density dynamic spectrum is shown on the right. {\it Middle: } A1 region overlaps an active region that produced a weak coronal jet (arrow). SPREDS for A1 reveals quasi-periodic type III bursts. {\it Bottom: }A2 overlaps an active region that produced a confined microflare (arrow). SPREDS for A2 reveals numerous coherent emission fine structures ($\delta t\lesssim 2\,s$; $\delta \nu \sim$10 -13\,MHz). Credits:~\cite{Atul19_typeIIIQPP_turb, Atul19_ARTB_microflare}}
   \label{fig:SPREDS}
   \vspace{-0.3cm}
\end{figure}
%It is hence important that, beyond the simple traditional type-casting of radio bursts, we need a systematic, robust classification of radio source flux and polarization variability for events of different activity types (e.g., jets, confined flares), energy scales (nano - major flares), and local physical parameters.
{ In addition, solar and stellar activity studies that rely primarily on total-flux dynamic spectra without snapshot spectroscopic imaging are susceptible to contamination from co-temporal bright events, as discussed in Sect.~\ref{sec:req_sol-stel}}.
%Owing to the multiwaveband imaging and magnetogram observations, 
{ For the Sun, we can combine} the coronal plasma diagnostics based on { EUV to X-ray imaging, magnetic field diagnostics based on vector magnetograms~\citep{Bobra14_SHRAPS}}, and non-thermal radio emission variability in a spatially resolved manner along the axes of intensity and polarization to { reliably constrain} the emission mechanism and { study} correlations between the radio source variability and local dynamics at the associated active regions. 

Figure~\ref{fig:SPREDS} illustrates the { use of spatially resolved dynamic spectra~\citep[SPREDS;][]{Atul17}, often also referred to as vector dynamic spectra~\citep{Chen18_typeIII_jet}, in separating the emission contributions from various co-temporal radio sources to the Sun-as-a-star dynamic spectrum.} 
The figure shows MWA radio contours of the Sun during Nov 3, 2014, revealing two bright radio sources (A1 and A2).
{ The zoomed-in images of the source regions at 94\AA\ from the Atmospheric Imaging Assembly~\citep[AIA;][]{Boerner12_AIAtempResp} onboard Solar Dynamic Observatory~\citep[SDO;][]{Pesnell12_SDO}} reveal their association with two different active regions. 
SPREDS for each radio source { was} made by recording the integrated flux density within an elliptical contour centered at the source for all spectroscopic snapshot images made with 0.5\,s and 160\,kHz averaging.
The A1 source showed signatures of quasi-periodic intermittent groups of type III bursts in its SPREDS, co-temporal with a coronal jet event. Meanwhile, the active region associated with the A2 source underwent a microflare. The A2 source was active even in the pre-flare phase with { QPPs in the radio emission} at $\sim$s and $\sim$10-15\,MHz scales. The coherent { flux density of A2 and the strength of QPPs} increased during the flare.
{ It is evident that the burst features in the Sun-as-a-star dynamic spectrum arise as a superposition of emissions from A1 and A2.}
The SPREDS for each source showed very different quasi-periodic spectro-temporal variability, which were analysed in detail along with multiwaveband data and magnetic field modeling in multiple publications~\citep{Atul19_typeIIIQPP_turb,Atul19_ARTB_microflare,atul21_structQPPs}.
These works could link the radio emission variability scales to physical scales at the active regions. 
%Such studies have also been done using other modern radio telescopes, like VLA and LOFAR~\citep[e.g.][]{Chen18_typeIII_jet,sharykin2018_LOFAR_dnn_withtypIIIb}.
{ Studies of this kind, spanning several events occurring in regions with varying physical parameters, will help characterize radio emission variability in flux density and polarization associated with different event types, and enable the development of robust physical diagnostics of the corresponding active regions.}

The angular resolution, superior sensitivity, and high fidelity imaging of SKA will enable a robust characterisation of true source morphology and its spectro-temporal flux and polarisation variability for even weak radio transients.
These studies will inform the { sensitivity and spectro-temporal resolution required} to track various types of { activity} in Sun-like stars.
Applying { solar-based insights and diagnostics to stellar dynamic spectra from the SKA telescopes could help infer the activity types and derive physical parameters of the unresolved active regions}.

\subsection{Generality of the standard flare model}\label{sec:stdflare_problem}
{ Despite its general success, several solar and stellar flare observations tend to question the applicability of the standard flare model. 
Given the diverse  
physical conditions that can exist at or around active regions, it is possible that certain events could deviate, at least partially, from the standard model pathway~\citep[e.g.][]{veronig02_neupertdev,Tristan23_Neupert-nonNeupert}.} 
%The question of possible degenerate physical pathways to produce the same observed multiwaveband flare light curves, as the Fig.~\ref{fig1}b template are also being discussed. 
{ Certain major} observations that { suggest the prevalence of non-standard processes beyond the standard flare model framework} are the missing radio burst problem and the departures from the Neupert effect. 

\subsubsection{The missing radio bursts problem}\label{missingradio}
{ %During a flare, the magnetic reconnection event generates energetic electron beams that produce various types of coherent bursts at different radio frequencies corresponding to the coronal heights they traverse (see, Fig.~\ref{fig2}).} 
As mentioned in Sect.~\ref{sec:emiss_mech}, coherent radio bursts that outshine the disk-integrated solar/stellar flux density are produced by accelerated electron beams, and are key diagnostics of flare-associated reconnection, CMEs, and EPEs. 
%The type I radio bursts that originate from small-scale magnetic reconnections in active regions without producing detectable SXR or EUV flares in Sun-as-a-star data~\citep{smith62_flare_typeI_correlStats,Elgaroy1970_typIcharacterisation,white2007_radiobursts,iwai14_typeI_finefeature_hists,Li17_typeI_smallEUVflare_link}. 
%Consequently, the various radio burst types (II, III, IV) serve as key diagnostics of space-weather drivers such as CMEs, SEPs, and major flare-associated particle acceleration.
Nevertheless, recent results from both stellar and solar observations appear to challenge this view.}

\paragraph{Stellar case.}
Young M dwarfs ($\lesssim$1 Gyr) possess strong, dynamic magnetic fields and are among the most magnetically active stars in the Galaxy~\citep{stepien94_Defn_Rx+Ro_Vs_activity_n_manyCorCurves,Barnes03_Rot_Vs_age_Vs_Activity,Donati09_Rev_Bfield,Vidotto14_B_Vs_age_n_rot}. They produce frequent large flares and superflares ($E>10^{33}$ erg) at rates of once every few days~\citep{dal20_flarestats,feinstein20_youngstarflarestats}.
Such flares are often associated with CMEs in the Sun, and are accompanied by coherent type II–IV bursts~\citep{McLean_solar_radiobook,Gopal11_PREconf,Miteva17_SEP-radburst_link}. Extensive searches have been carried out spanning hours to days on several active stars~\citep[e.g.][]{Osten2006,Osten2008,villadsen19_Cohbursts_but_notypeII} to detect solar-like burst types. 
{ Until now, a type IV has been detected in Proxima\,Cen~\citep[M5.5V;][]{Zic20_typeIV_ProximaCen}, a type IV and long-duration type III in AD\,Leo~\citep[M4V;][]{Atul24_ADLeotyIV}, and a type II in StKM 1-1262~\citep[M0V;][]{CTK25}.}
%Only recently were two type IV events and a possible type III detected from Proxima Cen and AD Leo~\citep{Zic20_typeIV_ProximaCen,Atul24_ADLeotyIV}, and a type II has now been detected on an M dwarf \citep{CTK25}.
While the { dearth of type II bursts in M dwarfs} could be attributed to high Alfvén speeds suppressing coronal shocks~\citep{Gomez22_stellarCMEprob}, the general lack of type III/IV or any flare-linked radio bursts in several strong stellar flares remains difficult to reconcile with the standard flare model~\citep[e.g.][]{2005A&A...436..241S,MacGregor18_proxima_flares}. 
%Based on the number of M-dwarfs that were scanned by the LOFAR Two-metre Sky survey~\citep{shimwell17_LoTSS} within the same distance as their type II source, \cite{CTK25} estimated a rate 10$^{-3}$ events per day per star of events with similar radio luminosity in M-dwarfs. 

\paragraph{Solar case.}
For the Sun, direct imaging and in-situ measurements of CMEs and SEPs allow detailed correlation studies between radio bursts and space-weather phenomena. %Nearly all type II bursts and $>85$\% of type IV bursts are CME-associated, with 75–80\% of type IVs linked to fast, wide, geo-effective CMEs~\citep{Gopal11_PREconf,Anshu21_CME_typeIV,anshu23_typeII_stats,Atul24_DHtypeIV}.
{ Despite the high association rates, well above 85 - 90\%, for type IV and II bursts with strong CMEs, only fewer than 5\% of CMEs produce such bursts~\citep{Anshu21_CME_typeIV,anshu23_typeII_stats,Atul24_DHtypeIV}.} 
%\cite{2007SoPh..240..263B} had shown that about 15\% of the solar SXR flares above GOES class C5 lacked any form of coherent radio bursts (type I - V) in 100 - 4000\,MHz and decimetric emissions. The authors showed that the sources of radio-quiet flares tend to be outside of |75|$^\circ$ source longitude, hinting at enhanced radio absorption towards the limb.
Certain studies on type IV burst detectibility in metric wavebands have suggested the role of radiation absorption by dense structures in the non-detection of type IV emissions~\citep[e.g.][]{Hillaris16_typeIV,Nasrin18_DHtypeIV_occult_streamerCMEshock}.
\citet{Atul24_DHtypeIV} compiled a catalog of type IV bursts in the 1 - 10\,MHz band using decades of data from Wind and STEREO spacecraft that observed the same flare/CME event from multiple vantage points simultaneously. The authors showed that type IV emission exhibits a strong inherent beaming effect. { Hence, the} spacecraft that observed the burst source within $\pm$60$^\circ$ line-of-sight always recorded the most intense event with { relatively greater} spectro-temporal extent. 
\citet{Atul24_typeIV-sun-ADLeo_comp} further noted that the favorable line-of-sight { orientation} of { AD Leo’s active latitudes} may have aided in detecting its type IV burst. 
{ However, neither the occultation/absorption nor the inherent beaming effect can fully explain the non-detection of type II/IV bursts in previous AD\,Leo monitoring studies~\citep{Osten2006,villadsen19_Cohbursts_but_notypeII}, and in many solar flares/CMEs.}

Another noteworthy finding in this context is that nearly 17\% of HXR solar flares in a sample of 201 events studied by \cite{benz05_Radio-quietHXRflares} had no type of associated coherent radio bursts in the 100 - 4000\,MHz band~\citep{2007SoPh..240..263B}, even though HXR emission is another diagnostic of non-thermal electrons. 
All 201 events studied were of SXR flare class greater than C5, which are relatively strong events capable of driving CMEs~\citep{yashiro06_flare-CME_corr,kazachenko23_ErupConfflare_ML}.
%{ High-dynamic range imaging of solar active regions is essential to detect and model even weak radio transients during X-ray/EUV flares. We also need sensitive long-duration multiwaveband monitoring studies of flare stars to explore optical/X-ray flares that lack radio emission enhancements. The superior imaging sensitivity of the SKA telescopes will enable both of these studies.}

\subsubsection{Departures from Neupert effect}\label{neupert}
In the standard flare model, magnetic reconnection not only drives electron beams upward into the corona, producing coherent bursts, but also downward { producing impulsive microwave bursts observable with the SKA-Mid and HXR flares as they impact the chromosphere (see, Fig.~\ref{fig1}b). Later, these energetic electrons dissipate energy heating the coronal loops, causing SXR flares. 
%Energetic electrons propagating downwards are expected to generate impulsive microwave flares in the GHz bands (see Fig.~\ref{fig1}b), probed by SKA mid.
The Neupert effect, which relates the time integral of impulsive-phase emission to the thermal flare light curve, is therefore a cornerstone of the standard flare model~\citep[see Sect.~\ref{sec:intro}][]{neupert68}.
Besides, the GBR relationship that applies universally across flares in main-sequence stars highlights a tight correlation between SXR and microwave (5-8\,GHz) flare peak flux densities.}
Despite this universal relationship between SXR and microwave flares, several { multiwaveband stellar flare observations} found cases where either SXR flares lacked radio counterparts or vice versa~\citep{Osten2005_100percVpol, 2005A&A...436..241S}. Stellar flare studies that { used U-band data to track impulsive chromospheric emissions also found cases where the Neupert effect seems to be violated}~\citep{Osten2005_100percVpol,Tristan23_Neupert-nonNeupert}.
Even in the Sun, statistical studies show that about 20 - 30\% of the flares depart from the expected Neupert effect~\citep[e.g.][]{1993SoPh..146..177D,2002A&A...392..699V}. For instance, \cite{Benz17_noHXR-microwave_SXRflare} reported a solar SXR flare which had no counterparts in HXR and 17\,GHz. In another study, \cite{2008AstL...34..704S} reported microwave bursts that lasted well into the SXR flare decay phase with high flux density, indicating possible multiple episodes of particle acceleration. Thus, cases of continued particle acceleration without { significant heating} and thermal flares without clear impulsive phase emissions { have been reported in many} solar flares.
%The missing microwave bursts reported in several SXR flares of GOES class >C5, mentioned in Sect.~\ref{missingradio}, also connect to this problem.

{ The SKA telescopes will be pivotal in tackling the problems of missing radio bursts and the deviations from the Neupert effect in the Sun and stars. High-dynamic range snapshot spectroscopic imaging of solar active regions enabled by the SKA telescopes will help detect and model even weak radio transients during X-ray/EUV flares in great detail. 
Additionally, these telescopes will enable sensitive long-duration multiwaveband monitoring studies of several cool stars, letting us identify and model the radio emission variability during optical/X-ray flares.
These studies will help explore the general applicability of the standard flare model, characterise the solar and stellar events that deviate from the model, and compare their possible mechanisms.}

\subsection{Plasma emission and low frequency stellar radio bursts}
With its spatial resolution and sensitivity, { the SKA-Low telescope} will be a superior instrument for the study of small-scale features on the Sun. { Though radio wave scattering} in the corona is expected to limit the instrument's spatial resolution { to an extent, the SKA-Low telescope will still advance the range
of observable spatial and temporal scales of variability in the solar
corona, probing the time and spatial scales of energy release and particle acceleration.}
As noted earlier, there has been some success in obtaining dynamic spectra of radio bursts for nearby flare stars, but the observations are sporadic and do not provide a complete picture. There are reasons to think that plasma emission will have different properties in other types of stars.
For example, free-free absorption decreases as the ambient temperature increases, so we expect that plasma emission will be detectable to higher frequencies than on the Sun, in stars with hotter
coronae~\citep[e.g.][]{White95_RSCVn_circPol}. { However,} coronae in some young active stars may be more extended and possess higher density than the Sun. { In such a case,} if electron beams are released low in the corona, they
might not be able to propagate as far as they can on the Sun, { leading to differences in the observed spectro-temporal variability in the radio dynamic spectrum compared to a solar flare of similar strength.
%Similarly, as discussed earlier, it may be more difficult to launch a CME on a star with much stronger and widespread coronal magnetic fields, providing a barrier to eruptions, in turn limiting the occurrence of type II-like frequency-drifting radio bursts.
Enhanced density in strong post-flare magnetic structures could also lead to significantly weak coherent emission flux density and impact the polarization signatures.}
SKA’s sensitivity will allow a deeper and more thorough census of the
occurrence of radio bursts on a wider sample of stellar types and ages than we can expect with
current instrumentation. { This will} allow us to better understand the nature of the coherent plasma emission process under varied physical scenarios { existing} in different stars, and contrast with the results from detailed solar imaging observations of various types of active phenomena.
\subsection{Linear polarization}
Recent confirmation of linear polarization in stellar ~\citep[e.g.][]{Zic20_typeIV_ProximaCen,bastian22_auroralemiss_UVCet} and solar \citep{DKO25} radio emission is a potentially important development that can be exploited to investigate
atmospheric structure. 
Conventional wisdom has always been that Faraday rotation in the solar
or stellar atmosphere is so large at microwave frequencies that it will wipe out linear polarization
even over narrow bandwidths, leaving just the circular polarization that is commonly observed and
well understood in the solar context. Understanding how linear polarization can survive despite
the known presence of a magnetized plasma atmosphere on active stars will likely reveal important
information about stellar atmospheric structure. At present, we only have a very limited picture of
the occurrence of linear polarization: we need to know much more about what types of stellar radio
emission (burst, flare, quiescent) and what types of stars do and do not exhibit linear polarization to understand the role of coronal structure. { The polarization sensitivity of the SKA telescopes} will be { pivotal} for
this field of study.
%Meanwhile, recent solar studies using MWA had also detected traces of linear polarisation in solar active regions. 
{ Sensitive} solar spectropolarimetry across the wide radio band will allow us to explore linear polarisation in various solar conditions and gain a better understanding of the underlying mechanism across different physical conditions existing in the Sun and stars.

\subsection{Evolution of stellar flaring activity}
Another paradigm in the solar-stellar field is that stars are born rapidly rotating and slow down as they age. { Magnetic activity declines in older stars, since it is tied to \Prot.} 
Observing the occurrence of stellar flares powered by magnetic activity is important, both for understanding the evolution of the star and because of the interest in the harshness of the space weather conditions close-in Earth-like exoplanets have to endure over the lifetime of different cool stellar types.
{ Addressing the latter question is crucial for identifying stellar populations within the $P_{\rm rot}$–age–$T_{\rm eff}$ parameter space that host relatively less hazardous environments for habitable-zone exoplanets and are therefore more likely to support biospheres.} 
%how long do activity levels inconsistent with life occur on
%different types of stars as they age? 
{ The sensitivity of the SKA telescopes} will allow us to measure the activity levels of
less active stars than we currently have access to, providing a much broader census of the evolution
of stellar activity with stellar type and age.

\subsection{Quiescent coronal activity diagnostics}
Non-flaring or quiescent radio emission offers a powerful diagnostic of persistent non-thermal activity in stellar coronae~\citep[e.g.][]{Bastian90_FlarestarsVLA,guedel02_Rev_stellarRadioEmiss,villadsen19_Cohbursts_but_notypeII}. 
%~\citep{ash2012_flarestats,klimchuk15_CorHeat_keyaspects, Sven16_ALMA_science,dal20_flarestats}.
{ Quiescent microwave emission, within 1 - 30\,GHz, is commonly detected from active stars, but its physical origin remains unclear. In principle, it could arise as thermal gyroresonance emission from a very hot corona with strong large scale magnetic field coverage}~\citep[e.g.][]{white94_radiodMconstraints}.
%Quiescent emission { in the microwave band} is commonly detected on active stars, and its origin is not well understood. In principle, it could be thermal gyroresonance emission from
%a very hot corona if the stellar surface has sufficient coverage of strong coronal magnetic fields~\citep[e.g.][]{white94_radiodMconstraints}. 
The associated spectral flux density is expected to increase with frequency.
{ However, the observed quiescent microwave spectra of flare stars are typically flat \citep[e.g.][]{guedel89_braodband_emissdMe,plant24_UVVet_Qemiss}, suggesting the prevalence of long-lived non-thermal coronal emission.}

%Exploring the characteristics of quiescent radio emission across the SKA’s wide spectral range will therefore yield valuable metrics of quiescent coronal activity across main-sequence stars as a function of rotation \Prot–age–\teff\ plane. The superior sensitivity of SKA will enable the detection of quiescent emission from numerous nearby cool stars out to distances of $\sim$50 pc with on-source integrations varying from a few minutes to under an hour, depending on distance and expected stellar flux density.
\begin{figure}[!htb]
 \vspace{-0.2cm}
  \centering
  \includegraphics[width=\textwidth,height=0.2\textheight]{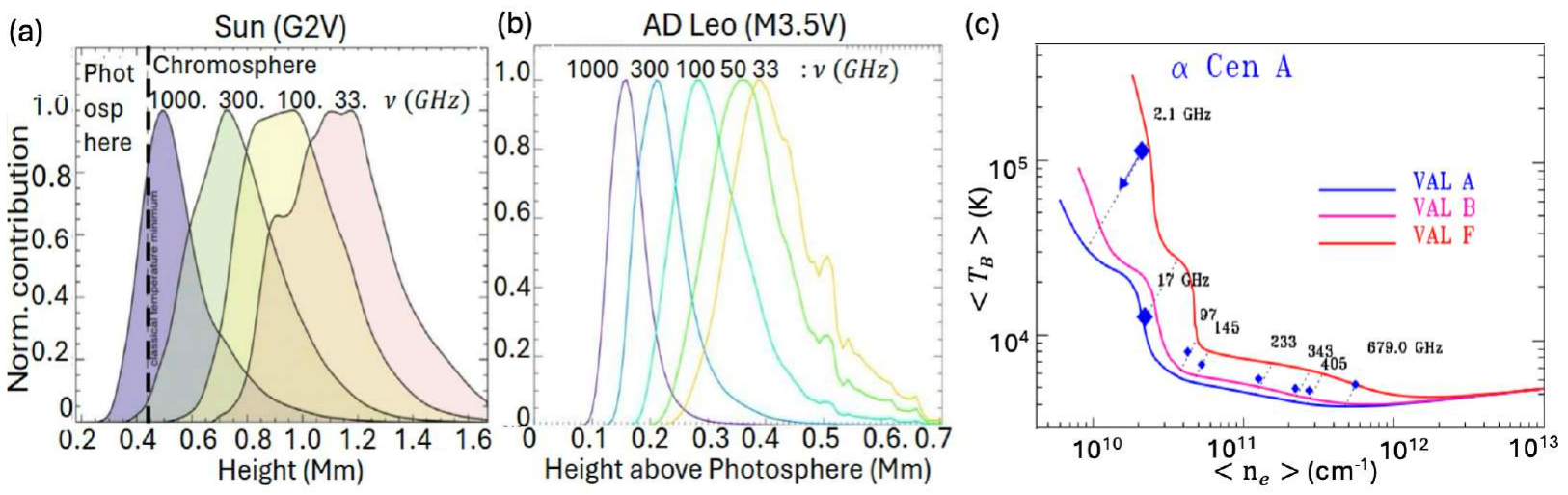}
\vspace{-0.8cm}
   \caption{{ Formation of mm emission. {\it (a - b): }Normalised contribution function for various mm frequencies computed using 3D atmopsheric models for Sun and AD\,Leo. {\it (c):} Scaled solar chromopsheric model fit to the observed mm-\Tbp\ of $\alpha$\,Cen\,A~\citep{2018MNRAS.481..217T}.}
   } 
   \label{fig:mm_orig}
   %\vspace{-0.8cm}
\end{figure}

\begin{figure}[!htb]
 \vspace{-0.2cm}
  \centering
  \includegraphics[width=\textwidth,height=0.2\textheight]{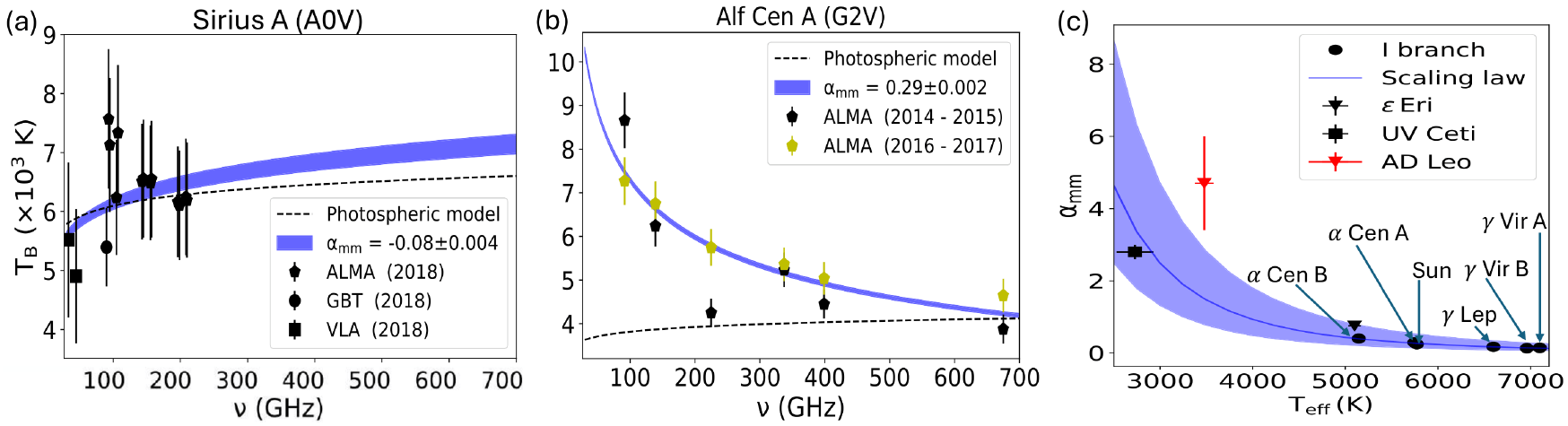}
\vspace{-0.8cm}
   \caption{{ Quiescent mm-\Tbp\ of main-sequence stars. {\it (a - b): }The mm-\Tbp, power-law spectral fit, and the photospheric model spectrum are for Sirius\,A and $\alpha$\,Cen\,A. {\it (c):} Scaling law between \alfmm\ and \teff\ for `I' branch cool stars, including Sun~\citep{Atul21_EMISSAI,Atul22_EMISSAII,Atul25_ADLeo_mm}. AD\,Leo, `C' branch star (\Prot = 2.23\,d, age$\sim$0.25\,Gyr, \teff $\sim$ 3500\,K) deviates from the trend.} 
   %{\it (b): } Scaled solar model fit to $\alpha$\,Cen\,A (G2V; \teff = 5745\,K, age $\sim$ 4.85\,Gyr, \Prot = 22\,d) data in 2 - 880\,GHz range~\citep{2018MNRAS.481..217T}. {\it (c-d):} \Tbp\ of $\alpha$\,Cen\,A and the Sun as a star along with powerlaw fits. The photospheric emission model from the PHOENIX code is overlaid. For the Sun, predictions from 1D atmospheric models (FAL A \& C) with a chromosphere are also provided. {\it (e): } The mm-\Tbp\, spectral power law fit and PHOENIX model prediction are provided for Sirius\,A - the hot A-type star that lacks chromospheric heating and hence shows \alfmm$<$0, as opposed to F - M dwarfs (see Panel a). The photospheric emission model from PHOENIX matches the data well. The bottom panel presents the fractional photospheric \TB\ based on observations, along with the expectation from the photospheric emission model (PHOENIX). The spectrum of the quiet and active Sun is provided for comparison~\citep{White20_MESAS}.
   } 
   \label{fig:alfmm}
   %\vspace{-0.8cm}
\end{figure}

%Combined with chromospheric activity diagnostics from ALMA and JVLA in bands $>$15\,GHz, SKA will enable the derivation of the complete sub-THz spectrum that will provide a tomographic view of steady state activity across the upper photosphere to corona.
Studies on quiescent stellar emission at frequencies $>$15\,GHz, using ALMA and JVLA, have yielded key insights into the steady heating of stellar chromospheres and transition regions~\citep[e.g.][]{Osten06_EVLac_QuiesFlare,Bastian18_EpsEri,2018MNRAS.481..217T,White20_MESAS,suresh20_EpsEri_RadioSEDmodel,Atul21_EMISSAI,Atul22_EMISSAII,plant24_UVVet_Qemiss,Atul25_ADLeo_mm}.
{ For instance, several studies have demonstrated thermal origin of the quiescent millimeter (mm) emission from stellar chromospheres, with longer wavelengths probing higher heights~\citep[e.g.][]{Selhorst14_NoRH_SSN_3GHz_srcCount,Sven16_ALMA_science,white04_solar-stellar_conn,2018MNRAS.481..217T}.
The mm brightness temperature (mm-\Tbp) spectrum thus acts as a linear thermometer probing the effective electron temperature across stellar chromospheric heights.
Figure~\ref{fig:mm_orig}a shows the normalised contribution functions for various mm frequencies derived using a 3D solar atmospheric model, highlighting the varying emission formation height ranges as a function of frequency. Panel b in the figure estimates the same for AD\,Leo, using a 3D atmospheric model for the star~\citep{Sven13_ADLeoChromModel}.
Figure~\ref{fig:mm_orig}c shows the results of a data-constrained modeling of the observed mm-\Tbp\ of $\alpha$Cen\,A. 
%using scaled versions of solar atmospheric models, VAL A, B, and F. 
The \Tbp\ values at various frequencies, including portions of the SKA-Mid's operational band, are consistent with the range predicted by scaled solar chromospheric thermal emission models (VAL A, B, and F).
%found to agree well with the range expected from the scaled solar chromospheric thermal emission models, VAL A, B, and F. 
%that the observed spectrum can be explained as originating from chromopsheric fits to the $\alpha$\,Cen\,A spectrum with a point at 17\,GHz close to the upper edge of the SKA mid band.

In a sample of ALMA-detected stars, \cite{Atul21_EMISSAI} showed that mm-\Tbp\ of F-M type stars with an active outer atmosphere deviated significantly from a purely photospheric model spectrum computed using the PHOENIX code given the stellar parameters~\citep{1999JCoAM.109...41H}.
Meanwhile, the A-type stars in the sample that lack active chromospheres agreed with the respective photospheric emission model.
Figure~\ref{fig:alfmm} shows the mm-\Tbp\ data points, the best-fit power-law model, and the PHOENIX-based photospheric model for an A and a G dwarf.
%Also, the lower the stellar \teff\ or equivalently higher the surface magnetic activity~\citep{Donati09_Rev_Bfield}, the higher the mm-\Tbp\ deviated from the respective photospheric model spectrum. 
%with lower \teff, which are expected to be more magnetically active.
In fact, the mm-\Tbp\ spectral index, \alfmm\ (\Tbp$\sim \nu^{-\alpha_{\rm mm}}$), which is related to the strength of chromospheric heating, scales with \teff~\citep{Atul21_EMISSAI,Atul25_ADLeo_mm}.}
%Besides, in a sample of slow rotating `I' branch stars, a power-law scaling has been derived between \alfmm\ (\Tbp$\sim \nu^{-\alpha_{\rm mm}}$) and \teff, making \alfmm\ a robust metric of quasi-steady chromospheric heating across the cool main sequence. }
%{ studies have demonstrated thermal origin of the quiescent mm emission from stellar chromopsheres, with longer wavelengths probing higher heights~\citep[e.g.][]{Selhorst14_NoRH_SSN_3GHz_srcCount,Sven16_ALMA_science,white04_solar-stellar_conn,2018MNRAS.481..217T}.} 
%\alfmm\ could thus serve as a metric of quasi-steady chromospheric heating and activity amongst stars across the cool main-sequence.}
%{In `I' branch stars that belong to a population of Sun-like, relatively low activity}.
Figure~\ref{fig:alfmm}d shows the scaling law between \alfmm\ and \teff,{ for a sample of stars belonging to the `I' branch activity population, like our Sun.} 
%comparing the Sun as a star alongside other main-sequence stars and highlights the data point from the active star, AD\,Leo, that clearly deviates. 
{ Clearly AD\,Leo, which is a fast-rotating `C' branch star, deviates from the trend}.
%Panel b shows the scaled solar model fits to the $\alpha$\,Cen\,A spectrum with a point at 17\,GHz close to the upper edge of the SKA mid band.
%Panels b-c show the mm-\Tbp\ data points, the best fit curve with the \alfmm\ mentioned, and the PHOENIX-based photospheric emission model.
%Clearly, the A-type star Sirius A agrees with the photospheric model \alfmm$<$0 unlike F-M type stars, and the lower \teff\ stars \alfmm\ steadily rises. Panels c and d show the power-law fits to mm-\Tbp\ derived from archival data for $\alpha$\,Cen\,A and the Sun-as-a-star, along with predictions from a photospheric 1D atmospheric emission model, derived using the PHOENIX code~\citep{1999JCoAM.109...41H}. For the Sun, predictions from solar chromospheric 1D models, FAL A and C, for quiet and active Sun~\citep{2004A&A...419..747L} match the observations. Figure~\ref{fig:alfmm}e shows the mm-\Tbp\ data for an A-type star, Sirius\,A, that matches well with the PHOENIX-based photospheric emission model. The \alfmm\ of Sirius\,A is also negative as opposed to cool stars in Fig.~\ref{fig:alfmm}a clearly separating stars with and without active hot outer atmospheres in the \alfmm\ - \teff\ plane~\citep{Atul22_EMISSAII}. The comparison of the solar spectrum with that of Sirius\,A also clearly shows the lack of atmospheric heating in the A-type star and how emission beyond 10\,GHz can provide useful diagnostics of chromospheric heating and activity across the main-sequence.

With the combined capabilities of SKA, ALMA, and JVLA, it will be possible to model the quiescent sub-THz spectrum for stars across the \Prot–age–\teff\ plane and to derive physically meaningful metrics of quasi-steady { heating} and activity across the corona to chromosphere, and { explore its correlations with stellar magnetic field and plasma parameters}.
SKA will be crucial in providing diagnostics in the { 5 - 15\,GHz range}. 
%Besides, surface magnetic field maps using Zeeman Doppler Imaging~\citep[e.g.][]{Donati09_Rev_Bfield,morin10_ZDI_late_dM*,Bellotti23_ZDIADLeo_Bevol} and near-simultaneous SXR-optical data for stars observed by the SKA, can further help constrain stellar quiescent atmospheric plasma parameters. 

\subsection{Evolution of the Sun and solar analogs in time}
Understanding how the activity of our Sun evolved over ages and will evolve into its future is fundamental to not only solar long-term activity research but also the question of the exposure of habitable zone planets (Venus, Earth, and Mars) to the evolving space weather in the past and into the next few gigayears.
The study of magnetic activity in solar-like stars { at} different evolutionary stages has advanced significantly due to { large surveys by various space ground-based high energy spectropolarimetric facilities}~\citep[e.g.][]{Marsden14_Bcool,vansaders16_weakBbrake-oldstars,davenport19_age-activityevol,feinstein20_youngstarflarestats}. The sensitive radio observations from SKA can provide valuable constraints on { magnetic field} and wind properties~\citep[e.g.][]{González_2002}.
Extending { SPREDS-based active region and space weather diagnostics} on long-term stellar flare datasets will help { infer the occurrence rates and properties of active events, especially those with potential space weather consequences}.
{ Performing such studies on long-term monitoring datasets from Sun-like stars of varying ages will be crucial in obtaining a statistical picture of activity evolution in a typical solar-like star during its lifetime~\citep{2020SSRv..216..143G}.}

\section{Conclusions}
%Magnetic activity in cool main-sequence stars shapes the space-weather environment experienced by nearby habitable-zone planets. Steady stellar coronal heating, particle acceleration, and winds drive persistent atmospheric erosion and chemical reactions in the atmospheres of nearby planets. Meanwhile, strong transient flares, energetic particle events, and coronal mass ejections can destabilize planetary atmospheres and significantly impact atmospheric chemistry.
{ The various kinds of magnetic activity in the cool main-sequence stars are often observed as enhanced emissions or flares in the radio to X-ray bands.
The Sun, with its unique spatially resolved observations and in-situ energetic particle measurements, provides a detailed view of how active regions evolve and produce flares, which cause major space weather impacts.}
%The physical framework and multiwavelength emission mechanisms in the widely accepted standard model of stellar activity are grounded primarily in solar observations. 
The wealth of multiwavelength solar imaging datasets serves as the benchmark for constructing disk-integrated Sun-as-a-star { diagnostics of active regions and space weather, extendable to stellar flare observations.}
{ Of the various flare signatures, radio bursts provide key diagnostics of magnetic reconnection and particle acceleration, which are the central ingredients of the standard flare model.}

The field of solar–stellar activity seeks to apply solar-based models and Sun-as-a-star diagnostics to interpret unresolved stellar active regions and to assess space-weather conditions during both active and quasi-steady emission phases. Comparative analyses of solar and stellar events also allow us to identify departures from the standard flare and activity framework, study the long-term evolution of magnetic activity in Sun-like stars, and investigate both quiescent and active atmospheric conditions across different stellar types. Since this young and rapidly evolving field depends on both solar and stellar observations, the advances enabled by the SKA telescopes in both domains will be highly beneficial to solar–stellar activity research.
%Characterizing magnetic activity requires a tomographic exploration tool, which the wideband SKA-Low and Mid arrays readily supply. 
The frequencies { probed by the SKA telescopes}, 0.05 - 15\,GHz will probe a variety of non-thermal particle acceleration processes { in the Sun and stars} across heights from low corona to interplanetary space. 
%The high-cadence spectro-polarimetry will enable well-constrained modeling of events, providing diagnostics of plasma and magnetic-field properties.
{ The high cadence, high dynamic range, spectro-polarimetric imaging of the Sun, enabled by the SKA telescopes, will allow the exploration of robust Sun-as-a-star diagnostics of active regions and space weather.}
%This is crucial to explore various types of activity without being impacted by co-temporal intense emissions in the image plane, and derive robust Sun-as-a-star diagnostics based on emission variability across flux density, frequency, time, and polarisation. The combination of high spectro-temporal resolution, excellent sensitivity, and high polarization purity will enable detailed identification and classification of emission variability linked to diverse active-region dynamics and space-weather phenomena. Together with co-temporal multiwavelength images, these data will support the development of a comprehensive template library for stellar applications.
On the stellar side, { the exceptional sensitivity and survey speed of the SKA telescopes} will enable high-cadence, spectro-polarimetric observations of stellar flare dynamic spectra { from several stars within several tens to a few hundred parsecs. 
These stellar datasets and the Sun-as-a-star diagnostics based on detailed solar imaging studies will facilitate various comparative solar stellar science cases. Some of these science cases are discussed in the chapter, namely, the test of the standard flare model across stellar systems, exploring radio emission mechanisms under varying stellar and solar atmospheric conditions, modeling quiescent atmospheric activity in the sun and stars, and exploring the evolution of activity and space weather conditions in stars across \Prot-age-\teff\ parameter space.}

\paragraph{Acknowledgements}
AM is partly supported by NASA’s STEREO project and the LWS program.
SW acknowledges support by the Research Council of Norway, project number 325491, through its Centres of Excellence scheme, project number 262622 (RoCS). AM acknowledges the helpful discussions with Dr. Tim Bastian.

\bibliographystyle{abbrvnat-maxbibnames4}
%\bibliography{chapter,references} % if your bibtex file is called example.bib
\bibliography{chapter_v1} % if your bibtex file is called example.bib

\end{document}